\documentclass[12pt]{article}
\usepackage{amssymb,amsmath,amsthm,amsfonts,amscd}
\newcommand\be{\begin{equation}}
\newcommand\ee{\end{equation}}
\newcommand\bea{\begin{eqnarray}}
\newcommand\eea{\end{eqnarray}}

\newcommand{\fatalpha}{{\bf \alpha \kern -0.44em \alpha}}
\newcommand{\fatsigma}{{\bf \sigma \kern -0.54em \sigma}}
\newcommand{\tpchi}{{\bf \chi \kern -0.35em \chi}}
\newcommand{\llambda}{{\bf \lambda \kern -0.45em \lambda}}

\bibliography{plain}
\title{\bf Geometry of Two-Qubit State and Intertwining Quaternionic Conformal Mapping Under Local Unitary Transformations}
\author{ G. Najarbashi $^{a,b}$
 \thanks{E-mail:najarbashi@tabrizu.ac.ir} ,
S. Ahadpour $^{a,b}$ \thanks{E-mail:ahadpour@uma.ac.ir} , M. A.
Fasihi $^{c}$ \thanks{E-mail:ma-fasihi@azaruniv.edu} ,
Y. Tavakoli $^{a}$\thanks{E-mail:tavakoli$\_$ phys@yahoo.com} \\
\\ $^a${\small Department of Theoretical Physics and Astrophysics,
Tabriz University, Tabriz 51664, Iran.} \\
$^b${\small Department of Physics, Mohaghegh Ardabili University,
Ardabil 56199-11367, Iran.}\\
$^c${\small Department of Physics, Azarbaijan University of
Tarbiat Moallem, Tabriz 53714-161, Iran.}} \pagebreak

\begin{document}
\maketitle
\newpage 

\begin{abstract}
In this paper  the geometry of two-qubit systems under local
unitary group $SO(2)\otimes SU(2)$ is discussed. It is shown that
the quaternionic conformal map intertwines between this local
unitary subgroup of $Sp(2)$ and the quaternionic M\"{o}bius
transformation which is rather a generalization of the results of
Lee et al (2002 Quantum Inf. Process. 1 129).


 {\bf PACs Index: 03.67.–a 03.65.Ud
 }
\end{abstract}

\newpage
\section{Introduction}
There has been considerable recent interest in understanding the
structure of one,two, three and multi-qubit systems, from the
geometrical point of view \cite{Lee,Mosseri,
Lev1,Lev2,Lev3,Bogdan,Heydari1,Heydari2}. The relation between
\emph{conformal map} (or Hopf fibration in \cite{Mosseri}) and
single qubit and two-qubit states have first been studied  by
Mosseri and  Dandoloff \cite{Mosseri} in quaternionic skew-field
and subsequently  have been  generalized to three-qubit state
based on octonions by Bernevig and Chen \cite{Bogdan}. Also  some
attempts have been made to figure out the notion of entanglement
and basic geometry of the space of states
\cite{Lev3,Heydari1,Heydari2}. From an information-theoretic
standpoint, the construction of well-defined entanglement measure
typically relies on the concept of entanglement monotone which is
non-increasing under local operations and classical communication.
Such transformations are called LOCC \cite{Lev2,Horo1,Horo2}. For
instance, the most widely utilized measure for two-qubit, is
concurrence introduced by Wootters \cite{Wootters}.
\par
However it seems that  there is also another geometrical approach
to describe pure two-qubit states called conformal groups
\cite{Di-franc}. As is typical in physics, the local properties
are more immediately useful than the global properties, and the
local unitary transformation  is of great importance. Therefore in
this paper we pursue a different approach to study the geometrical
structure of two-qubit states under local unitary subgroup of
$Sp(2)$ \cite{Leo2}. We show that the quaternionic conformal map
(QCM) of a pure two-qubit system intertwines between the local
subgroup $Sp(2)$  and corresponding \emph{quaternionic M\"{o}bius
transformations} (QMT) \cite{Asla,Harvey,Porter} which can be
useful in theoretical physics such as quaternionic quantum
mechanics \cite{Adler}, quantum conformal field theory
\cite{Di-franc,Ginsparg} and quaternionic  computations
\cite{Fernandez}. However the action of transformations that
involve with non-commutative  quaternionic skew-field on a spinor
(living in quaternionic Hilbert spaces) is more complicated than
the complex one. Roughly speaking one must distinguish between the
left and right actions of a quaternionic transformations on a
given state (e.g see  \cite{Leo1}). This anomalous property of
quaternionic transformation lead us to define the special QMT.
\par
The
 paper is organized as follows: In Section 2. we briefly summarize
one-qubit geometry and conformal map in a commutative diagram. In section 3, we introduce the
basic geometric structure together with basic background material, incorporating all the information
we need for characterization of two-qubit geometry.
Section 4 devoted to study the commutativity of QCM in details. The paper is ended with a brief conclusion and
one appendix.
\section{One-qubit geometry}
We will denote by ${\mathcal{H}}_{d}^{\mathbb{F}}$  the Hilbert
space of dimension $d$ in  $\mathbb{F}=\mathbb{C}$ or
$\mathbb{Q}$. Let us consider an arbitrary one-qubit pure state in
complex two dimensional Hilbert space
${\mathcal{H}}_{2}^{\mathbb{C}}$
\begin{equation}
|\psi\rangle=\alpha_{1}|0\rangle+\alpha_{2}|1\rangle\quad\quad,\quad
|\alpha_{1}|^2+|\alpha_{2}|^2=1\quad\quad,\quad
\alpha_{1},\alpha_{2}\in \mathbb{C}.
\end{equation}
We summarize the results of Ref.\cite{Lee} in a  commutative
diagram fashion convenient for our purposes as:
\[\begin{CD}
{\mathcal{H}}_{2}^{\mathbb{C}}         @>{\mathcal{P}}>>      \mathbb{\widetilde{C}} \\
@V{A}VV               @VV{\mathcal{F}_{_{A}}}V \\
{\mathcal{H}}_{2}^{\mathbb{C}}         @>>{\mathcal{P}}>
\mathbb{\widetilde{C}}
\end{CD}\]
where $\mathcal{P}$ is conformal mapping for one-qubit system,
i.e.,
\begin{equation}
\mathcal{P}(|\psi\rangle):=\alpha_{1}\alpha_{2}^{-1}\in
\mathbb{\widetilde{C}}= \mathbb{C} \cup \{\infty\},
\end{equation}
and
 $\mathcal{F}_{_{A}}\in PSU(2)=SU(2)/\{\pm I\}$ is M\"{o}bius transformation corresponding to
$2\times2$ matrix $A\in SU(2)$
\begin{equation}
A=\left(\begin{array}{cc} a& b
\\c&d
\end{array}\right)\ \longleftrightarrow\ \mathcal{F}_{_{A}}(z)=\frac{az+b}{cz+d}\quad\quad a,b,c,d,z \in
\mathbb{C}.
\end{equation}
The M\"{o}bius transformations generate the conformal group in the
plane and  can be identified using stereographic projection with
conformal transformations on the sphere. The action of the
M\"{o}bius group on the Riemann sphere is transitive in the sense
that there is a unique M\"{o}bius transformation which takes any
three distinct points on the Riemann sphere to any other set of
three distinct points. Commutativity of the above diagram means
that for any one-qubit state $|\psi\rangle$ and any $A\in SU(2)$
we have
\begin{equation}
\mathcal{F}_{_{A}}\mathcal{P}(|\psi\rangle)=\mathcal{P}(A|\psi\rangle).
\end{equation}
This shows that the conformal mapping $\mathcal{P}$ intertwines
between any single qubit unitary operation $A$ and its
corresponding M\"{o}bius transformation $\mathcal{F}_{_{A}}$. It
is tempting to try to extend this diagram to the system of
bipartite two-qubit systems. However due to the deference between
the dimensions of single qubit and two-qubit  systems, all the
above processes must be modified in a convenient way which is the
task of the next section.
\section{Basic tools and definitions}
We will require some preliminary definitions and results.
Therefore this section devoted to provide some basic tools and
background to attack to the geometrical properties of two-qubit
pure states.
\subsection{Quaternionic conformal map}
The Hilbert space ${\mathcal{H}}_{4}^{\mathbb{C}}$ for the
compound system is the tensor product of the individual Hilbert
spaces
${\mathcal{H}}_{2}^{\mathbb{C}}\otimes{\mathcal{H}}_{2}^{\mathbb{C}}$
with a direct product basis
$\{|00\rangle,|01\rangle,|10\rangle,|11\rangle\}$. A two-qubit
pure state reads
\begin{equation}\label{twoqubit}
|\psi\rangle=\alpha|00\rangle+\beta|01\rangle+
\gamma|10\rangle+\delta|11\rangle\quad\quad\quad
\alpha,\beta,\gamma,\delta,\in \mathbb{C},
\end{equation}
with normalization condition
$|\alpha|^2+|\beta|^{2}+|\gamma|^{2}+|\delta|^{2}=1$. Using
quaternionic skew-field $\mathbb{Q}$ we can equivalently restate
every $|\psi\rangle\in {\mathcal{H}}_{4}^{\mathbb{C}}$
 by a quaterbit $|\widetilde{\psi}\rangle\in
{\mathcal{H}}_{2}^{\mathbb{Q}}$ as \cite{Mosseri}
\begin{equation}
\mathcal{Q}(|\psi\rangle):=|\tilde{\psi}\rangle=q_{1}|\tilde{0}\rangle+q_{2}|\tilde{1}\rangle\quad,\quad
q_{1}=\alpha+\beta \textbf{j}\quad,\quad q_{2}=\gamma+\delta
\textbf{j}\quad,\quad |q_{1}|^2+|q_{2}|^2=1.
\end{equation}
One can easily check  that the map $\mathcal{Q}$ is a complex linear map that is
$$
\mathcal{Q}(c_{1}|\psi_{1}\rangle+c_{2}|\psi_{2}\rangle)=
c_{1}\mathcal{Q}(|\psi_{1}\rangle)+c_{2}\mathcal{Q}(|\psi_{2}\rangle)\quad\quad\forall
\ c_{1},c_{2}\in \mathbb{C}.
$$
The simplest way to introduce  conformal map for two-qubit system
is to proceed along the same line as for one-qubit case, but using
quaternions instead of complex numbers (see appendix):
\begin{equation}\label{confq}
\mathcal{P}(|\tilde{\psi}\rangle):=q_{1}q^{-1}_{2}=\frac{1}{|q_{_{2}}|^2}[(\alpha+\beta
\textbf{j})(\overline{\gamma}-\delta
\textbf{j})]=\frac{1}{|q_{2}|^2 }(S+C\textbf{j}) \in
\mathbb{\widetilde{Q}}= \mathbb{Q} \cup \{\infty\},
\end{equation}
where the Schmidt $(S)$ and concurrence $(C)$ terms are defined as follows
\begin{equation}
S:=\alpha\overline{\gamma}+\beta\overline{\delta}\quad\quad,\quad
C:=\beta\gamma-\alpha\delta.
\end{equation}
It should be  mentioned that the map $\mathcal{P}$ is related to
the projection of the second Hopf fibration of the form
\begin{equation}\label{}
    \mathcal{P}: \mathbb{Q}^2 \longrightarrow \mathbb{Q}P^1
\end{equation}
where $\mathbb{Q}P^1$, is the one dimensional quaternionic
projective space.
 Note that if $S=0$ then $|\psi\rangle$ has Schmidt
decomposition
\begin{equation}
|\psi\rangle=|q_{1}||0\rangle_{1}
|e\rangle_{2}+|q_{2}||1\rangle_{1}|f\rangle_{2},
\end{equation}
where $\{|e\rangle,|f\rangle\}$ is two orthonormal basis for
second qubit. Moreover $C$ is proportional to one of entanglement
measure
$\mathcal{C}(|\psi\rangle):=\langle\psi|\sigma_{y}\otimes\sigma_{y}|\overline{\psi}\rangle
$ called \emph{concurrence} \cite{Wootters} where
$\overline{\psi}$ denotes the complex conjugation and $\sigma_{y}$
is one of Pauli spin operators. Concurrence is widely used to
quantify entanglement of two-qubit systems. In fact
$2\overline{C}=\mathcal{C}$ and if $\mathcal{C}=0$ then
$|\psi\rangle$ unentangled in the sense that it can be written as
a tensor product of two pure state of individual subsystems, i.e.,
$|\psi\rangle=|\phi\rangle_{1}|\varphi\rangle_{2}$. The
quaternionic conformal map $\mathcal{P}'$ defined by
$\mathcal{P}':=q^{-1}_{2}q_{1}$ is distinct from $\mathcal{P}$ and
may be interpreted as one for  dual space. Indeed it can be easily
 verify that
$\mathcal{P}'(\langle\tilde{\psi}|)=\overline{\mathcal{P}(|\tilde{\psi}\rangle)}$.
\subsection{Local unitary subgroup of $Sp(2)$}
Before discussing the local unitary subgroup of $Sp(2)$ we would
add some short discussion on the transformation properties of the
two-qubit entangled state and its quaternionic representative. It
would simplify later presentation considerably if we represent the
$|\psi\rangle$ of Eq. (\ref{twoqubit}) by the $2\times 2$ matrix
\begin{equation}\label{}
    \Psi=\left(\begin{array}{cc}
       \alpha & \beta \\
       \gamma & \delta \\
     \end{array}\right)
\end{equation}
which gives rise to the transformation property
$\vert\psi\rangle\to A'\otimes A\vert\psi\rangle$, for $A',A\in
SU(2)$
\begin{equation}\label{matrixform}
    \Psi\mapsto A'\Psi A^{T}
\end{equation}
where $A^{T}$ refers to the transposed of $A$.  The quaternionic
version of this transformation is
\begin{equation}\label{quatact}
    \vert\tilde{\psi}\rangle\to A'\vert\tilde{\psi}\rangle (a-\overline{b}\textbf{j}).
\end{equation}
\par
The quaternionic counterpart of group
$SU(2)$ for two-qubit system seems to be  group $Sp(2)$ which is
defined as
\begin{equation}
Sp(2):= \{B\in GL(2,\mathbb{Q}) : B^{\dagger}B=I\},
\end{equation}
or equivalently  can be expressed by
\begin{equation}
Sp(2):= \{U\in U(4) : UJU^T=J\}.
\end{equation}
where $J:=I\otimes {\varepsilon}$ , with $\varepsilon\equiv
-i{\sigma}_{2}$.
 Since the two-qubits systems have entanglement property
therefore we will consider operations which do not change the
entanglement measure (concurrence) throughout the diagram. As it
is well known such operations must act locally on each individual
qubit . Therefore we restrict ourself  to local subgroup
$\mathcal{B}\simeq SO(2)\otimes SU(2)$ of group $Sp(2)$ where its
corresponding complex form $\mathbb{C}B$ reads
\begin{equation}\label{CB}
\mathbb{C}B=\left(\begin{array}{cc}
 \cos\theta&
\sin\theta\\-\sin\theta &\ \cos\theta
\end{array}\right)\otimes\left(\begin{array}{cc}
 a&
 b\\-\bar{b} &\bar{a}
\end{array}\right),
\end{equation}
and we will investigate the problem for this local unitary
operations in the next section.  Within this interesting scenario
it is a trivial and well-known fact that the measure of
entanglement (concurrence) does not change regarding the local
unitary transformations like $\mathcal{B}$.
\subsection{Quaternionic M\"{o}bius transformations}
The main difficulties in establishing the quaternionic approach to
M\"{o}bius transformation is the non-commutativity of the
quaternions. Beside that, it is rather seamless to carry over much
of the complex theory. For $M\in SL(2,\mathbb{Q})$ we define the
QMT
\begin{equation}\label{QMT}
M=\left(\begin{array}{cc} m_{11}& m_{12}
\\m_{21}&m_{22}
\end{array}\right)\ \longleftrightarrow\ \mathcal{F}_{_{M}}(q):=(q \ m_{11}+m_{12})(q \ m_{21}+m_{22})^{-1}
\quad\quad m_{ij} \in \mathbb{Q},
\end{equation}
with the conventions $\mathcal{F}_{M}(\infty)=m_{11}m_{21}^{-1}$ and $\mathcal{F}_{M}(-m_{22}m_{21}^{-1})=\infty$.
As is the case with $\tilde{\mathbb{C}}$, there is a M\"{o}bius transformation taking any three given points to
any other three points, however, it is not unique.
It is easily seen that $\mathcal{F}_{_{MM'}}=\mathcal{F}_{_{M}}\circ \mathcal{F}_{_{M'}}$ where
matrix multiplication is defined in usual way, i.e.,
$ MM'=(m_{i1}m'_{1j}+m_{i2}m'_{2j})$.
The set of all such transformations forms a group under composition.
This set is identified naturally with the quotient space $PSL(2,\mathbb{Q})=SL(2,\mathbb{Q})/\{\pm I\}$.
 However, again due to the non-commutativity of the field
$\mathbb{Q}$, in addition to $\mathcal{F}_{_{M}}$ there are
several possibilities to define the QMT \cite{Porter}, i.e.,
$$
\begin{array}{c}
  \mathcal{F}'_{_{M}}(q)=(q\ m_{21}+m_{22})^{-1}(q \ m_{11}+m_{12})  \\
  \mathcal{F}''_{_{M}}(q)=(m_{11}\ q+m_{12})(m_{21} \ q+m_{22})^{-1} \\
  \vdots  \\
\end{array}
$$
Therefore the extending of the commuting QCM which intertwines
between the group $\mathcal{B}$ and the corresponding QMT fixing
the measure of entanglement is our purpose. As we will see in the
next section to attribute physical interpretation for the QMT and
to fit the problem in a commutative setting it is necessary to
choose Eq.(\ref{QMT}) as a preferable definition of QMT. This
choice for the QMT  is based on the implicit fact that we treat
the space of quaternionic spinors as a right module
(multiplication by scalars from the right).
\section{Two-qubit geometry}
We now proceed one step further, and investigate the results of
the previous section for  two-qubit pure states under the action
of local unitary subgroup $\mathcal{B}$ of $Sp(2)$.\\
In this section it will be shown that a direct substitution of the
definitions of previous section leads to the commutativity of the
following diagram
\[\begin{CD}
{\mathcal{H}}_{2\otimes2}^{\mathbb{C}}         @>{\mathcal{Q}}>>      {\mathcal{H}}_{2}^{\mathbb{Q}}    @>{\mathcal{P}}>> \mathbb{\widetilde{Q}} \\
@V{\mathbb{C}B}VV               @VV{B}V           @VV{\mathcal{F}_{_{B}}}V\\
{\mathcal{H}}_{2\otimes2}^{\mathbb{C}}       @>{\mathcal{Q}}>>
{\mathcal{H}}_{2}^{\mathbb{Q}}    @>{\mathcal{P}}>>
\mathbb{\widetilde{Q}}
\end{CD}\]
The purpose of the diagram is to verify that wether the QCM
intertwines between the operator $B\in \mathcal{B}$ and the
corresponding QMT $\mathcal {F}_{_{B}}$. This  implies that for
any two-qubit pure state we expect that the following equalities
\begin{equation}\label{eq}
\mathcal{P}\mathcal{Q}(\mathbb{C}B|\psi\rangle)
=^{?}\mathcal{P}B(\mathcal{Q}|\psi\rangle)=^{?}\mathcal{F}_{_{B}}\mathcal{P}(\mathcal{Q}|\psi\rangle),
\end{equation}
hold for any $|\psi\rangle\in {\mathcal{H}}_{4}^{\mathbb{C}}$. By
choosing of each equality above one can breakdown this diagram
into three pairs of commutative pieces. Therefore we study each of
them which  every two-qubit (quaterbit) can be influenced by the
maps introduced above. The above equalities follow from the
following calculations.
\subsection{Calculating $\mathcal{P}\mathcal{Q}(\mathbb{C}B|\psi\rangle)$}
It is convenient to start with the first statement
in Eq.(\ref{eq})
\begin{equation}\label{path1}
\mathcal{P}\mathcal{Q}(\mathbb{C}B|\psi\rangle)=\mathcal{P}(|\tilde{\psi}'\rangle)=
\mathcal{P}(q'_{1}|\tilde{0}\rangle+q'_{2}|\tilde{1}\rangle) =
q_{1}'q_{2}'^{-1},
\end{equation}
where $q'_{1}$ and $q'_{2}$ are results of the action of Eq.
(\ref{CB}) on the general  two-qubit pure state Eq.
(\ref{twoqubit}) followed by the map $\mathcal{Q}$ as
$$
q'_{1}=\alpha'+\beta'\textbf{j}=[( a \alpha+b \beta) \cos \theta +  (a \gamma+b \delta)\sin \theta]+
[(\bar{a}\beta -\bar{b}\alpha)\cos\theta+(\bar{a}\delta-\bar{b}\gamma)\sin\theta]\textbf{j}\ ,
$$
\begin{equation}\label{q12}
    q'_{2}=\gamma'+\delta'\textbf{j}=[( a \gamma+b \delta) \cos \theta
-( a \alpha+b \beta)\sin \theta]+ [(\bar{a}\delta
-\bar{b}\gamma)\cos\theta-(\bar{a}\beta-\bar{b}\alpha)\sin\theta]\textbf{j}\
.
\end{equation}
On the other hand the Eq. (\ref{path1}) can be expressed in terms
of Schmidt and concurrence terms
\begin{equation}\label{SC1}
\mathcal{P}\mathcal{Q}(\mathbb{C}B|\psi\rangle)=\frac{1}{|q'_{2}|^2
}(S'+C'\textbf{j}),
\end{equation}
where the norm of $q'_{2}$ is
$$
|q'_{2}|^2=|q_{2}|^2\cos^2 \theta + |q_{1}|^2
\sin^2\theta-\sin2\theta \ \rm{Re }(S),
$$
and the $S'$ and $C'$  are given by
\begin{equation}\label{Schm}
S'=\alpha'\overline{\gamma'}+\beta'\overline{\delta'}=
\cos^2 \theta \ S-\sin^2\theta\ \bar{S}+\frac{1}{2}\sin2\theta(|q_{2}|^2-|q_{1}|^2)\ ,
\end{equation}
\begin{equation}\label{conc}
C'=\beta'\gamma'-\alpha'\delta'=C.
\end{equation}
We observe that independent of the parameters $a,b$ and  $\theta$,  the concurrence term $C'$
is invariant under the action of $\mathbb{C}B$.
This observation is well known and fulfills our expectations that entanglement can
be changed only by global  transformations.
An interesting situation arises when  $\theta=0$  e.g., $S'=S$. This case
is coincide with the results of  Mosseri et
al in \cite{Mosseri}.\\
This is not the only way to get above results dealing with the
geometry of two-qubit states. In the next subsection we shall
establish  the similar operations on a quaterbit and find the same
results.
\subsection{Calculating $\mathcal{P}B(\mathcal{Q}|\psi\rangle)$}
Considering the diagram we can proceed another approach to
understand more about the two-qubit entangled state. Unlike in the
definition of ${\cal B}$ in order to correctly represent the
complex transformation on the two-qubit state, the separable
$Sp(2)$ transformation on the quaternionic spinor should be
represented by left action of the $2\times 2$ matrix  $A'\in
SO(2)$ containing $\sin\theta$ and $\cos\theta$, and right
multiplication with the quaternion $a-\overline{b}\textbf{j}$, as
in Eq. (\ref{quatact}). Hence by
 applying the $B\in \mathcal{B}$ on a quaterbit
$|\tilde{\psi}\rangle$ one can get
$$
 \left(\begin{array}{cc}
 \cos\theta&
\sin\theta\\-\sin\theta &\ \cos\theta
\end{array}\right)\vert\tilde{\psi}\rangle
(a-\overline{b}\textbf{j})=\left(\begin{array}{c}
                    q_{1}' \\
                    q_{2}' \\
                  \end{array}\right),
$$
where $q_{1}'$ and  $q_{2}'$ are the same as in Eq. (\ref{q12}).
Therefore the relevant part of the crucial diagram (the first
quadrangle) is commutative, i.e.,
\begin{equation}\label{quadrangle}
   \mathcal{Q}(\mathbb{C}B|\psi\rangle)
=B(\mathcal{Q}|\psi\rangle).
\end{equation}
 It is clear that applying the QCM on the
both side of the above equation lead to the first equality in Eq.
(\ref{eq}).
\subsection{Calculating $\mathcal{F}_{_{B}}\mathcal{P}(\mathcal{Q}|\psi\rangle)$}
We have already shown that the first equality in Eq. (\ref{eq})
holds. Let us now see what happen on a quaterbit regarding the
action of QMT. Using the linear map $\mathcal{Q}$ together with
QCM on a  two-qubit pure state in Eq. (\ref{twoqubit}) yield
\begin{equation}
\mathcal{P}\mathcal{Q}(|\psi\rangle)=\mathcal{P}(|\tilde{\psi}\rangle)=
\mathcal{P}(q_{1}|\tilde{0}\rangle+q_{2}|\tilde{1}\rangle)
= q_{1}q_{2}^{-1}=\frac{1}{|q_{2}|^2}(S+C\textbf{j})
\end{equation}
Furthermore this point is  mapped under the action of the QMT in Eq. (\ref{QMT}) as follows
$$
\mathcal{F}_{_{B}}\left(\frac{1}{|q_{2}|^2}(S+C\textbf{j})\right)=\left(\frac{1}{|q_{2}|^2}(S+C\textbf{j})
 (a-b\textbf{j})\cos\theta+(a-b\textbf{j})\sin\theta\right)
$$
$$
 \hspace{4.5cm}\times\left(\frac{1}{|q_{2}|^2}(S+C\textbf{j})
 (-a+b\textbf{j})\sin\theta+(a-b\textbf{j})\cos\theta\right)^{-1}
$$
$$
 \hspace{4.8cm}=\frac{
\cos^2 \theta \ S-\sin^2\theta\
\bar{S}+\sin\theta\cos\theta(|q_{2}|^2-|q_{1}|^2)\ + C\ \textbf{j}
}{|q_{2}|^2\cos^2 \theta + |q_{1}|^2 \sin^2\theta-\sin2\theta \
\rm{Re }(S)}
$$
Again we get precisely the same result as the two previous
subsections meaning that the second equality in (\ref{eq}) holds.
This in turn implies that  the QCM intertwines between the
operator $B\in \mathcal{B}$ and the corresponding QMT $\mathcal
{F}_{_{B}}$. This is what we have expected to see. To sum up we
have the three commutative diagrams for  two-qubit pure states as
mentioned above.
\subsection{ $SU(2)\otimes SO(2)$ transformation}
So far we have considered the $SO(2)\otimes SU(2)$ subgroup of
$Sp(2)$ and find the total commutative diagram. Let us now see
what has been gained in considering the separable subgroup
$\mathcal{B}'\simeq SU(2)\otimes SO(2)$, in the sense that $SU(2)$
and $SO(2)$ act on the first and second particles of the pure
two-qubit state respectively. It is easy to show that for this to
occur, one must consider the map
$$
M\longrightarrow \mathbb{C}M: M=\left(\begin{array}{cc} m_{11}&
m_{12}
\\m_{21}&m_{22}
\end{array}\right)\ \longrightarrow
 \left(\begin{array}{cc|cc}
   z_{11}^{(1)} & z_{12}^{(1)} & -\bar{z_{11}}^{(2)} & -\bar{z_{12}}^{(2)} \\
   z_{21}^{(1)} & z_{22}^{(1)} & -\bar{z_{21}}^{(2)} & -\bar{z_{22}}^{(2)} \\ \hline
    z_{11}^{(2)} & z_{12}^{(2)} & \bar{z_{11}}^{(1)} & \bar{z_{12}}^{(1)} \\
   z_{21}^{(2)} & z_{22}^{(2)}& \bar{z_{21}}^{(1)} & \bar{z_{22}}^{(1)} \\
  \end{array}\right),
$$
where $ m_{ij}=z_{ij}^{(1)}+z_{ij}^{(2)}\textbf{j}\ $ and
$z_{ij}^{(1)},z_{ij}^{(2)}\in \mathbb{C}$, which in turn induces
the following definition for $2\times 2$ symplectic group
\begin{equation}
Sp(2):= \{U\in U(4) : U^T J'U=J'\}.
\end{equation}
 where in this case $J'= \varepsilon \otimes I $. Note that
in the transformation  $A'\otimes A|\psi\rangle$ and subsequently
in its matrix form Eq. (\ref{matrixform}), $A'$ and $A$ could be
any member of group $SU(2)$. On the other hand in the quternionic
version of transformation $SO(2)\otimes SU(2)$ on a quaterbit we
were not worry about left or right action of the $A'\in SO(2)$ on
a quaternionic spinor. However  here in our discussion $A'\in
SU(2)$ while $A\in SO(2)$ and the former acts
 on the quaternionic spinor. Therefore unlike the
$SO(2)\otimes SU(2)$ case, one must distinguish between the left
and right actions  regarding quaternionic version of the separable
subgroup $SU(2)\otimes SO(2)$ on a  quaterbit
$|\widetilde{\psi}\rangle\in {\mathcal{H}}_{2}^{\mathbb{Q}}$.
Roughly speaking in this case we must use the left action as
follows
$$
 \left(\begin{array}{c}
                    a q_{1}+b q_{2} \\
                    -\bar{b} q_{1}+\bar{a} q_{2} \\
                  \end{array}\right)
(\cos\theta-\sin\theta \textbf{j})=\left(\begin{array}{c}
                    q_{1}' \\
                    q_{2}' \\
                  \end{array}\right)
$$
which implies that the first quadrangle in the  main diagram is
commutative, i.e.,
\begin{equation}\label{}
   \mathcal{Q}(\mathbb{C}B'|\psi\rangle)
=B'(\mathcal{Q}|\psi\rangle),
\end{equation}
where $B'\in \mathcal{B}'$. But unfortunately, in this case there
is no apparent way to pick a particular QMT  in order to get total
commutative diagram and hence the problem of intertwining QCM will
cease to exist.
\section{Conclusion}
In this paper we considered the action of $SO(2)\otimes SU(2)$
part of quaternionic group $Sp(2)$ on a  two-qubit pure state as a
local transformation which obviously leaves invariant the measure
of entanglement. We have shown that QCM intertwines between local
unitary subgroup $Sp(2)$ and QMT. It is  rather interesting that
the three way are so well related to the important ingredients of
a pure two-qubit state which are Schmidt and concurrence terms. In
this investigation we found that other definitions for QMT do not
work. Another simple consequence of our findings is that the
choice of $SU(2)$ action on the first particle leads to the some
essential changes  on the main diagram in the sense that just the
first quadrangle become commutative together with the fact that
one have to use the left action on the quaterbit. Moreover in
using $SU(2)\otimes SO(2)$ on the pure two-qubit state, there is
no QMT which make the diagram total commutative and subsequently
there is nothing to do with QCM.
\\

\textbf{Appendix: Quaternion}\\
The quaternion skew-field $\mathbb{Q}$ is an associative algebra
of rank 4 over $\mathbb{R}$ whose every element can be written as
$$
q=x_{0}+x_{1}\textbf{i}+x_{2}\textbf{j}+x_{3}\textbf{k}\quad\quad,\quad
x_{0},x_{1},x_{2},x_{3}\in \mathbb{R}\quad \mathrm{with} \quad
\textbf{i}^2=\textbf{j}^2=\textbf{k}^2=\textbf{ijk}=-1.
$$
It can also be defined equivalently, using the complex numbers
$z_{1}=x_{0}+x_{1}\textbf{i}$  and
 $z_{2}=x_{2}+x_{3}\textbf{i} $ in the form  $q=z_{1}+z_{2}\textbf{j} $
 endowed with an involutory antiautomorphism (conjugation) such as
$$
q=z_{1}+z_{2}\textbf{j} \in \mathbb{C}\oplus\mathbb{C}\textbf{j}\
\longrightarrow \
\bar{q}=x_{0}-x_{1}\textbf{i}-x_{2}\textbf{j}-x_{3}\textbf{k}=\bar{z}_{1}-z_{2}\textbf{j}.
$$
Every non-zero quaternion is invertible, and the unique inverse is
given by $q^{-1}=\frac{1}{|q|^2}\bar{q}$ where the quaternionic
norm $|q|$ is defined by $ |q|^2=q\bar{q}=|z_{1}|^2+|z_{2}|^2.$
The norm of two quaternions $q$ and $p$ satisfies $ |qp| = |pq| =
|p||q|.$
 Note that quaternion multiplication is non-commutative so that
 $\overline{q_{1}q_{2}}={\overline{q_{2}}}\ {\overline{q_{1}}}$ and $
\textbf{j}z=\bar{z}\textbf{j},$ where the last relation  have been
used in this
paper extensively.\\
On the other hand a two dimensional quaternionic vector space $V$
defines a four dimensional complex vector space $\mathbb{C}V$ by "
forgetting" scalar multiplication by non-complex quaternions
(i.e., those involving $\textbf{j}$ or $\textbf{k}$). Roughly
speaking if $V$ has quaternionic dimension $2$, with basis
$\{|\widetilde{0}\rangle,|\widetilde{1}\rangle\}$, then
$\mathbb{C}V$  has complex dimension 4, with basis
$\{|00\rangle,|01\rangle,|10\rangle,|11\rangle\}$. Moreover each
matrix $ M\in M( 2, \mathbb{Q})$, i.e., each linear map $
M=(m_{ij}):V\longrightarrow V$
 defines a linear map $\mathbb{C}M:\mathbb{C}V\longrightarrow \mathbb{C}V$
 i.e., a matrix $\mathbb{C}M\in M( 4, \mathbb{C})$. Concretely, in passing
from $V$ to $\mathbb{C}V$ each entry $
 m_{ij}=z_{ij}^{(1)}+z_{ij}^{(2)}\textbf{j}\ $
is replaced by $2\times2$ complex matrix which means that the map
$$
M\longrightarrow \mathbb{C}M: M=\left(\begin{array}{cc} m_{11}&
m_{12}
\\m_{21}&m_{22}
\end{array}\right)\ \longrightarrow
 \left(\begin{array}{cc|cc}
   z_{11}^{(1)} & -z_{11}^{(2)} & z_{12}^{(1)} & -z_{12}^{(2)} \\
   \bar{z_{11}}^{(2)} & \bar{z_{11}}^{(1)} & \bar{z_{12}}^{(2)} & \bar{z_{12}}^{(1)} \\ \hline
    z_{21}^{(1)} & -z_{21}^{(2)} & z_{22}^{(1)} & -z_{22}^{(2)} \\
   \bar{z_{21}}^{(2)} & \bar{z_{21}}^{(1)}& \bar{z_{22}}^{(2)} & \bar{z_{22}}^{(1)} \\
  \end{array}\right),
$$
is injective and it preserves the algebraic structures such as
$\mathbb{C}(M+M')=\mathbb{C}M+\mathbb{C}M' $ ,
$\mathbb{C}(MM')=(\mathbb{C}M)(\mathbb{C}M')$ and
$\mathbb{C}(M^\dagger)=(\mathbb{C}M)^\dagger$, where
$(M^\dagger)_{ij}=\bar{m_{ji}}$. It is easy to see that
$$
M(2,\mathbb{Q})=\{M\in M(4,\mathbb{C}) : JMJ^{-1}=\bar{M}\},
$$
where metric $J$ is given by
$$
J:=\left(\begin{array}{cccc}
  0 & -1 & 0 & 0 \\
  1 & 0 & 0 & 0 \\
  0 & 0 & 0 & -1 \\
  0 & 0 & 1 & 0 \\
\end{array}\right).
$$
Considering $GL(2,\mathbb{Q})\subseteq M(2,\mathbb{Q})$ for the
subset of invertible matrices; it is well known
\cite{Asla,Harvey,Leo2} that $M$ has a two-sided inverse in
$M(2,\mathbb{Q})$ if and only if the $
\mathbb{C}(M(2,\mathbb{Q}))$ is invertible in $M(4,\mathbb{C})$
which implies that $ \mathbb{C}(M(2,\mathbb{Q}))$ belongs to the
group $GL(4,\mathbb{C})$ which consisting of  all
 invertible $4 \times 4$ matrices.
Of course, in this description, we also have
$$
GL(2,\mathbb{Q})=\{M\in GL(4,\mathbb{C}) :\ JMJ^{-1}=\bar{M}\},
$$
$$
SL(2,\mathbb{Q})=\{M\in GL(4,\mathbb{C}) :\ \det M =1\ ,\
JMJ^{-1}=\bar{M}\}.
$$

\textbf{Acknowledgments}\\
It is a pleasure for one of the authors (GN) to acknowledge
enlightening discussions about quaternion conformal map with H.
Fakhri. We also acknowledge private communications with R.
Sufiani for which we are deeply grateful.
The authors also acknowledge the support from the Mohaghegh Ardabili University.


\end{document}